\documentstyle[preprint,aps]{revtex}
\def\be {\begin{eqnarray}}
\def\ee {\end{eqnarray}}
\def\beq {\begin{equation}}
\def\eeq {\end{equation}}

\def\bi {\begin{itemize}}
\def\ei {\end{itemize}}
\def\ben {\begin{enumerate}}
\def\een {\end{enumerate}}
\begin{document}
\draft
\preprint{NORDITA-95/8 A/N, SUNY-NTG-94-34, KUNS 1313}
\title{
On Neutrino Emission From Dense Matter\\
Containing Meson Condensates
}
\author{
V. Thorsson$^a$, M. Prakash$^b$, T. Tatsumi$^c$ and
C. J. Pethick$^{a,d,e}$
}
\address{
$^{a}$ NORDITA, Blegdamsvej 17, DK-2100 Copenhagen \O, Denmark \\
$^{b}$ Department of Physics, SUNY at Stony Brook, Stony Brook, NY 11974,
USA\\
$^{c}$ Department of Physics, Kyoto University, Kyoto 606-01, Japan\\
$^{d}$ Department of Physics, University of Illinois at Urbana-Champaign,\\
1110 West Green St., Urbana, IL 61801-3080, USA\\
$^{e}$Institute for Nuclear Theory, Physics-Astronomy Building, NK-12,\\
University of Washington, Seattle, WA 98195, USA
}

\date{\today}
\maketitle
\begin{abstract}
We consider the rate at which energy is emitted by neutrinos from the dense
interior of  neutron stars containing a Bose condensate of pions or kaons.
The rates obtained  are larger, by a factor of
2, than those found earlier, and are consistent with those
found for the direct Urca processes.
\end{abstract}


\narrowtext

\renewcommand{\thefootnote}{\arabic{footnote}}

It has been suggested that matter at several times
the density of normal nuclei,
as found
in the cores of collapsing massive stars or in the dense
interior of neutron stars,
may contain a
Bose condensate of pions~\cite{picollect} or kaons~\cite{kcollect}.
A number of aspects of meson condensation under such conditions
have been explored, one being the possible role played by condensates in
the thermal evolution of neutron stars.
Neutron stars are born with internal thermal energies of
some tens of MeV.  In the first $10^5 - 10^6$ years after
formation,  the chief mechanism for energy loss is the emission
of neutrinos and anti-neutrinos from matter in the interior of the
neutron star.
Which neutrino emitting reactions are most effective
in removing energy from the neutron star is the subject of
numerous studies. (For a review, see Ref.~\cite{cooling}.)
In 1965, Bahcall and Wolf~\cite{bw} showed that if pions were found in the
interior of neutron stars, the neutrino emission rate could be
considerably higher than that found in the absence of pions.
Following the suggestion that attractive p-wave interactions
could lead to pion condensation~\cite{picollect},
Maxwell, Brown, Campbell, Dashen and Manassah~\cite{mbcdm} (MBCDM),
computed the rate of neutrino energy emission
from matter containing a pion condensate.
Shortly after the idea arose that kaons could condense in matter due to
attractive s-wave interactions~\cite{kcollect}, Brown, Kubodera, Page and
Pizzochero~\cite{bkpp} and Tatsumi~\cite{tats} pointed out that kaon
condensation could also lead to rapid cooling, by a reaction analogous to that
considered for pions.

In this note, we point out what we believe is an error in the emission
rate originally evaluated by MBCDM and in nearly all subsequent studies.
The corrected
emission rates are consistent with those obtained recently in a separate study
of a closely related emission process, the direct Urca process~\cite{lpph}.
Correcting this error is expected to have little effect on  the general
discussion of the relative importance of meson induced cooling processes in
relation to other processes considered viable~\cite{cooling,lvpp}.

Briefly, the discrepancy between the rates found in the presence of
meson condensates on the one hand~\cite{mbcdm,bkpp,tats,fujii}
and the direct Urca process~\cite{lpph}, on the other, is apparent by
observing that the net rate of emission from matter with a meson
condensate should, in the limit of a vanishing condensate,
reduce to the rate for the direct Urca process for nucleons.
That it does not  may be seen as follows.
Consider the reactions
\be
n(\pi) \rightarrow n(\pi) + e^- + \overline{\nu}_e \,\,{\rm and} \,\,
n(\pi) + e^- \rightarrow n(\pi) + \nu_e  \,\,,
\label{picool}
\ee
\noindent
treated in MBCDM.
In Eq.~(\ref{picool}), $n(\pi)$ denotes an excitation which is a superposition
of a neutron and a proton, and reduces to a free neutron in the
absence of a pion
condensate.  The pion condensate is parametrized by a momentum $k$, and an
angle $\theta$ describing the strength of the condensate, or the degree to
which the pion is chirally rotated away  from its value in the
vacuum by the presence of matter~\cite{picollect}.  In the case of kaon
condensation, the condensate is spatially uniform, characterized only by an
angle $\theta$, describing the degree of  V-spin rotation of the kaonic ground
state~\cite{kcollect}. Several reactions have been considered in the presence
of a kaon condensate, and are listed in Table 1.
The pairs of reactions (1) and (2) of Table 1
are analogous to reactions (\ref{picool}) above,
while reactions (3) of Table 1, to be described below,
are related to the direct Urca process.
In all cases,
the rate of energy emission due to neutrinos
is given by the square of the relevant matrix
element $\cal M$, summed over over
spin states of all participants in the reaction:
\be
|{\cal M}|^2 = H^{\mu \nu} L_{\mu \nu} \,\,\,,
\label{m2}
\ee
\noindent
where $H^{\mu \nu}$ and $L_{\mu \nu}$, in covariant
form,
are the hadronic and leptonic contributions, respectively.
It is easy to check that Refs.~\cite{mbcdm,bkpp,tats,fujii}
are internally
consistent in the sense that differences reside only in the hadronic
contribution $H^{\mu \nu}$, and that the overall coefficient in
the emission rate is consistent
with the final result of MBCDM, Eq.~(53).
(The actual evaluation of emission rates requires rather involved phase space
integration and will not be carried out here.   For details, we refer to
the original references~\cite{mbcdm,bkpp,tats}.
We have checked that phase space integrals
have been correctly performed in the references quoted.  Therefore, we
confine the
discussion below to matrix elements only.)
Furthermore, by taking the $\theta \rightarrow 0$ limit in
Eq.~(4.15) of ref.~\cite{fujii},
the rate obtained for reaction (3) of Table 1,
one obtains a prediction for the
energy emission rate due to the neutron direct Urca process.
The emission rate for the direct Urca process thus obtained is a factor of 2
smaller than the recent result  of Ref.~\cite{lpph},  a non-relativistic
calculation that is easy to  check against the rate of neutron beta decay in
free space.

The direct Urca process for nucleons refers collectively to neutron beta decay
\be
n \rightarrow p + e^- + \overline{\nu}_e  \,\,\,\,,
\label{durca1}
\ee
and electron capture on protons
\be
p + e^- \rightarrow n + \nu_e  \,\,\,\,.
\label{durca2}
\ee
For many years,
it was thought that the
direct Urca process was forbidden
at low temperatures, momentum conservation never being
possible due to the relatively small proton and electron Fermi momenta
in relation to that of the neutrons.
However, Lattimer, Pethick, Prakash and Haensel~\cite{lpph}
( see also Boguta~\cite{boguta} ), recently
drew attention to the fact that in several recent models of
nuclear matter at high
density, the concentration of protons may indeed be so large that
these kinematical conditions may be met.  This could lead to
very rapid cooling of neutron stars, with cooling rates even higher
than those found in the presence of meson condensates.

The origin of the factor of 2 discussed above,
which we believe is an error arising
in the original work of MBCDM, is not immediately clear.  One possible
explanation is simply that the overall numerical coefficient in
the final emission rate, Eq.~(53), was underestimated.
(Note that MBCDM also stress the importance of
maintaining consistency with the neutron beta decay rate.) However, we
believe the problem already arises at an earlier stage of the calculation.
Preceding Eq.~(29) of MBCDM it is stated that the relevant matrix element,
$|{\cal M}|^2$, includes sums over spins of all participants in the reaction.
The
leptonic contribution $L^{\mu \nu}$ is obtained from Eq.~(146.3) of Lifshitz
and Pitaevskii~\cite{llrqt}.  However, the $L^{\mu\nu}$ of Ref.~\cite{llrqt}
is expressed in terms of
density matrices, which do  not include a sum over the spins of the electrons
in the final state, and would therefore seem to be inconsistent with the
definition preceding Eq.~(29) of MBCDM.  Summing over the spin states gives
the conventional electron projection operator, which is a factor of 2 larger
than the density matrix.  We therefore expect the
leptonic tensor, relevant for the emission of an
electron of four-momentum $p_e$, and
an antineutrino of four-momentum, $p_{\nu}$, to read
\be
L^{\mu\alpha} = 8 \, ( p_e^\mu p_{\bar\nu}^\alpha
+ p_e^\alpha p_{\bar\nu}^\mu
- (p_e \cdot p_{\bar\nu})g^{\mu\alpha}
- i \varepsilon^{\mu\alpha\gamma\delta}(p_e)_\gamma(p_{\bar\nu})_\delta)
\,\,\,,
\ee
which is a factor of 2 larger than
the corresponding expression given in Eq.~(30) of MBCDM.
Conversely, if we choose to use the density matrix
formulation, we need to sum over the final spins of the electrons,  as in the
example of Ref.~\cite{llrqt}, thereby obtaining an additional factor of 2 over
that given in MBCDM.

In what follows, we summarize the various emission rates discussed in
the literature with the required increase by a factor of 2.
It is most convenient to express the results in terms of the
emission rate for the closely related direct Urca process, Eqs.(\ref{durca1})
and (\ref{durca2}).
The rate at which energy is emitted per unit volume
by reactions (\ref{durca1}) and (\ref{durca2}),
in matter in which all the
participating fermions are degenerate, is~\cite{lpph}
\be
\dot{E}_{Urca} &=& \frac {457\pi}{10080}
\frac{ G_F^2 \cos^2\theta_C (1+3g_A^2) }{ \hbar^{10} c^5 }
m_n m_p \mu_e (k_B T)^6
\nonumber \\
&=& 2.21 \times 10^{26}~
\left(\mu_e \over {100 {\rm MeV}} \right) (1+3g_A^2) \cos^2\theta_C T_9^6
{}~{\rm {erg~cm^{-3}~sec^{-1}} }
\,\,\,.
\label{durcarate}
\ee
\noindent Here,
$G_F=1.436 \times 10^{-49} \,\,{\rm erg} \,\,{\rm cm}^3$ is the Fermi coupling
constant,
$\theta_C \simeq 0.223 $ is the Cabibbo angle,
$g_A=1.26$ is the nucleon axial--vector coupling constant,
$m_n$ is the neutron mass,
$m_p$ is the proton mass,
$\mu_e$ is the electron chemical potential,
and $T$ is the temperature,
$T_9$ being its value in units of $10^9$ K.
In Eq.~(\ref{durcarate}), we have neglected the in-medium modification of the
weak-interaction matrix  elements  and  the effects of possible superfluidity
of neutrons and superconductivity  of protons~\cite{super}. The energy emission
rate is $2 \dot{E}_{Urca}$  if the muon Urca process can also occur.

The energy emission rate of the reactions in Eq.~(\ref{picool})
is, for small values of $\theta$ (see MBCDM),
\be
\dot{E}_\pi =
g\left(\frac{p_e}{k}, \frac{p_n}{k} \right)
\frac{ \theta^2 }{4}
\left[ 1 + \left(\frac{ g_A  k}{ p_e} \right)^2 \right]  \dot{E}_{Urca}
\,\,\,,
\ee
where $p_e$ and $p_n$ are, respectively, the electron and neutron Fermi
momenta, and
\be
g\left(\frac{p_e}{k}, \frac{p_n}{k} \right) =
-\frac18 \sum (-1)^{n_{-}}
\left|
\pm 1
\pm \frac{p_e}{k}
\pm \frac{p_n}{k}
\pm \frac{p_n}{k}
\right|
\,\,\,.
\label{g}
\ee
\noindent
In Eq.~(\ref{g}), the sum is over the 16 ways of assigning signs
in the expression in the modulus, and
$n_{-}$ is the number of negative signs in a particular term.
The function $g$ goes to unity in the limit $k \rightarrow 0$.
When $p_e$ is less than the length of any momentum
vector that can be made up from $k$ and two momenta with magnitudes equal to
the neutron Fermi momenta, $g=p_e/k$\cite{mbcdm}.

In connection with pion condensation, we remark that if the proton
concentration is sufficiently large, it will be possible for the proton analogs
of the processes (\ref{picool}) to occur.
To the best of our knowledge, this process has
not been considered previously.  For this process to be allowed, the sum of two
proton Fermi momenta and the electron Fermi momentum must exceed $k$.  If this
process occurs, its rate will be comparable to that of the neutron process
(\ref{picool}).

In Table 1, we list the energy emission rate, in
terms of ${\dot{E}}_{Urca}$, for the possible reactions
in the presence of a kaon condensate described earlier.\footnote{ The treatment
of the axial current matrix elements here is consistent with that of
Ref.~\cite{bkpp},
and relates to that of Refs.~\cite{tats,fujii} by the replacement  $g_A
\rightarrow 1 $ in the matrix element of $A_8^\mu$.}
It is worth emphasizing that the reactions discussed above take  place only if
energy-momentum conservation can be satisfied among  the participants for the
temperatures and densities under  consideration.   In most models describing
conditions in the interior of cooling neutron  stars, energy-momentum
conservation is rather easily met in reactions (1)
and (2) of Table 1 (an important factor in their being considered as  a viable
cooling agent~\cite{mbcdm,bkpp,tats}), whereas  reactions (3) of Table
1 take place only if the concentration  of protons in the matter is
sufficiently large~\cite{tpl,fujii}.

In conclusion, we believe that we have resolved a discrepancy between the
published neutrino energy emission rates from  dense matter with condensates,
on
the one hand~\cite{mbcdm,bkpp,tats,fujii},  and that found for the direct Urca
process~\cite{lpph}, on the other.  Rates found earlier in the presence of a
condensate are correct if multiplied by a factor of 2. (The recent
Ref.~\cite{tpl} contains  this correction.)  We do not expect this correction
to significantly alter our current understanding of the thermal  evolution of
neutron stars~\cite{lvpp}.

The research of M. Prakash was supported in part
by the U.~S. Department of Energy under
grant DE-FG02-88ER40388, and that of C. J. Pethick by
the U.~S. National Science Foundation under grant
NSF AST93-15133 and by NASA under grant NAGW-1583.
T. Tatsumi is grateful
to T. Muto and H. Fujii for stimulating
discussions, and to NORDITA for supporting his participation in
the Workshop on Meson in Nuclei and Meson Condensation
in Copenhagen in April 1994.
One of us \linebreak
( C. J. P. ) is grateful to the Institute for Nuclear
Theory at the University of Washington for its hospitality
and the Department of Energy for partial support during
the completion of this work.

\begin{table}
\caption{Neutrino emissivity in the presence of a charged kaon condensate
in terms of ${\dot{E}}_{Urca}$, the neutrino emissivity
for the direct Urca process,
{}~Eq.~(5).  Process 1 is the
one considered in the original work of Brown, Kubodera,
Page and Pizzochero
{}~[6] and Tatsumi
{}~[7] .}
\begin{tabular}{lcr}

& Reaction &  ${\dot{E}}/{\dot{E}}_{Urca}$  \\  \hline

1
&
$ n(K) \rightarrow n(K) + e^- + \overline{\nu}_e $
& \\
& $ n(K) + e^- \rightarrow n(K) + \nu_e $
&
$ \frac 14 \sin^2\theta \tan^2\theta_C   $ \\

&&\\

2
&
$ p(K) \rightarrow p(K) + e^- + \overline{\nu}_e $
& \\
& $ p(K) + e^- \rightarrow p(K) + \nu_e $
& $ \sin^2\theta \tan^2\theta_C   $ \\

&&\\

3
&
$ n(K) \leftrightarrow p(K) + e^- + \overline{\nu}_e $
& \\
&
$ p(K) + e^- \rightarrow n(K) + \nu_e $
&
$ \cos^2(\theta/2)    $
\end{tabular}
\end{table}

\end{document}